%% file: rwArXivFinal.tex
\documentclass[twocolumn,laps,pra,amsmath,amssymb]{revtex4}

\usepackage[usenames,dvipsnames]{xcolor}
\usepackage{soul}
\input{defs}

\newtheorem{criterion}[theorem]{Criterion}
\newtheorem{principle}[theorem]{Principle}

\definecolor{nblue}{rgb}{0.3,0.3,1.0}
\definecolor{ngreen}{rgb}{0.2,0.7,0.2}
\definecolor{nred}{rgb}{0.9,0.1,0}

\newcommand{\ea}{{\it et al.}}
\newcommand{\beq}{\begin{equation}}
\newcommand{\eeq}{\end{equation}}
\newcommand{\erf}[1]{Eq.~(\ref{#1})}

\newcommand{\imbh}{inherently measurement-based}
\newcommand{\imb}{inherently measurement based}
\newcommand{\osmb}{only superficially measurement-based}
\newcommand{\rhodf}{{\rho}}

\begin{document}

\title{Discord in relation to resource states for measurement-based quantum computation}
\author{Eleanor G. Rieffel${}^1$ and Howard M. Wiseman${}^2$}
\affiliation{${}^1$ QuAIL, NASA Ames Research Center, Moffett Field, CA 94035 \\
${}^2$Centre for Quantum Computation and Communication Technology (Australian Research Council), 
Centre for Quantum Dynamics, Griffith University, Brisbane, QLD 4111, Australia}
\date{}

\begin{abstract}
We consider the issue of what should count as a resource for 
measurement-based quantum computation (MBQC). While a state that
supports universal quantum computation clearly should be considered 
a resource, universality should not be necessary given the existence of
interesting, but less computationally-powerful, classes of MBQCs. 
Here, we propose minimal criteria for a state to be considered
a resource state for MBQC. Using these criteria, 
we explain why discord-free states cannot be resources for MBQC, 
contrary to recent claims [Hoban \ea,~arXiv:1304.2667v1]. 
Independently of our criteria, we also show that the arguments of Hoban \ea, 
if correct, would imply that Shor's algorithm (for example) 
can be implemented by measuring discord-free states. 
\end{abstract}
\date{\today}
\maketitle

\section{Introduction}
\label{sec:intro}

In measurement-based quantum computation (MBQC), computation is carried
out by performing local measurements on an initial state. 
Classical processing of the results may take place in parallel, 
enabling the measurements to be performed adaptively. 
The foundational proposal \cite{Raussendorf+Briegel-01}
for MBQC introduced the class of cluster states for the 
initial states, but since then many other types of entangled 
states have been found that support universal quantum computation
\cite{vanDenNest06,Gross07,Briegel09}.
Universality is obviously a sufficient condition for a state to
be a resource state for MBQC, but seems too strong as a necessary 
condition given the recent discovery of classes of MBQCs such as {\it IQP}
\cite{Bremner11}. This class includes only non-adaptive MBQCs, and so is 
unlikely to be universal, but nevertheless appears to give
an advantage over classical computing.
While it is generally believed that some entanglement 
(but not too much! \cite{Gross09,Bremner09,Low09}) is required for
a state to be a resource for MBQC, that has yet to be established. 

Recently, in their paper 
``Exact sampling and entanglement-free resources for measurement-based 
quantum computation,''
Hoban \ea~\cite{Hoban13} challenged the notion that entanglement is necessary 
by proposing that there are resource states for certain types of MBQC 
that not only are entanglement free, but even discord free. 
The zero-discord claim is particularly surprising because discord-free 
states are ``essentially classical'' \cite{Hoban13}, as they comprise 
mixtures of locally orthogonal states. Prompted by their 
work, we address the question of how resource states for MBQC should 
be defined. We propose some minimal
criteria, and come to conclusions contrary to those of Hoban \ea\
In particular, we argue that discord-free states {\em cannot} be resource 
states for MBQC. We leave open the question of whether other 
entanglement-free states could be resource states for MBQC. 

Our paper is structured as follows. 
First, we propose the principle that for a state to be a resource state 
for MBQC, it must  support  computations that are \imb\ 
(Sec.~\ref{sec:keyPrinciple}). 
In Sec.~\ref{sec:simpleCriteria}, we give minimal (i.e., weak)
necessary conditions for a set of MBQCs to be \imb. 
We use these criteria in Sec.~\ref{sec:discOfHoban}  
to consider the work of Hoban \ea, 
and explain why their arguments do not support their conclusion. 
We show that if one accepted their arguments, then one 
would also have to believe that Shor's algorithm (for example)
could be implemented in MBQC using discord-free resource states. 
In Sec.~\ref{sec:genArg}, we propose a slightly stronger necessary condition, 
from which it follows that it would never make sense to claim that discord-free
states are resources for MBQC. Finally, we summarize our findings. 
Readers interested only in our discussion of Hoban \ea\ 
could go straight to Sec.~\ref{sec:discOfHoban}.  Readers interested in the 
much broader question raised by our title should study Secs.~II, III, and V. 

\section{A key principle}
\label{sec:keyPrinciple}

Throughout, we consider the standard model of MBQC \cite{Raussendorf+Briegel-01,RPbook} 
in which single-qubit 
measurements on a $\nu$-qubit quantum state $\rho$, together with
 a classical computational
processor, are all that is needed to carry out the computation. 
In general, the measurements are adaptive \cite{WisemanMilburn,Raussendorf02}
(that is, the choice of measurement 
basis depends on the results of previous measurements), 
 but some interesting MBQC algorithms, such as MBQC implementations 
of  {\it IQP} computations \cite{Browne11}, are non-adaptive.

To construct any argument that some given state is 
(or is not) a resource for MBQC, one must 
start with a sufficient (or necessary, respectively)  
criterion for the concept of ``resource for MBQC''.  
Prior to this work, necessary criteria for a quantum state to be a
resource for MBQC have not appeared in the literature, to our knowledge. 

The term ``resource'' is used most frequently 
when discussing which states support universal MBQC and which do not
(e.g., \cite{vanDenNest06,Gross07,Briegel09,Gross09,Bremner09,Low09}). 
Universality is 
certainly a sufficient condition, but seems too strong as a necessary 
condition for a state to be a resource for MBQC in light of results 
such as the following: (i) Anders and Browne \cite{Anders09} showed that 
single-qubit measurements on GHZ states can boost the extremely 
limited classical computational class $\oplus L$ (parity-$L$) to $P$. 
(ii) Bremner \ea\ \cite{Bremner11} showed  that {\it IQP} contains 
computations not in $P$ (unless the polynomial
hierarchy collapses to the third level) while unlikely to give
universal quantum computation. Here {\it IQP} is the class of 
``instantaneous quantum computations,'' that is,  those that 
can be carried out with non-adaptive measurements in the 
MBQC model. (iii) Hoban \ea~\cite{Hoban13} extended the last result to 
the class {\it IQP}$^*$.

At the other extreme, since any quantum state can be subjected to
single-qubit measurements, one might say that every quantum state
should be considered a resource for MBQC. On reflection, however, 
it is clear that this is too weak 
as a sufficient condition; it would remove any possibility of 
scientific advance on the question of resource states for MBQC. 
That the proper concept 
of a resource state for MBQC ought to require something 
between these two extremes should be a point of 
agreement between all researchers in the field. 

One could try to characterize  which resources support (or do not support)
computationally interesting MBQC. The problem is
that, almost certainly, 
not all of the interesting MBQC tasks are known. 
Instead, we concentrate on a weaker property, namely,
what makes a set of  MBQCs worthy of the name MBQC. 
That is, we introduce the 
notion of {\it \imbh\ computations}, and give a series of necessary 
conditions for sets of MBQCs to be considered inherently 
measurement-based. 

We do not attempt to redefine MBQC itself to include only inherently
measurement-based quantum computations. 
The reason is that this would be an impediment to discussing MBQC, 
just as excluding from quantum computations ones 
that are not inherently quantum would be an impediment to discussing 
quantum computation. By not using such exclusive language, the community 
is able to make useful statements such as ``all reversible 
classical computations can be translated into a quantum computation 
of precisely the same efficiency.'' The concept of ``inherently quantum'' 
is useful, just as is the concept of ``\imb'' (as we will see), 
but in neither case is it fruitful to exclude computations 
that are not inherently of the type under consideration. 

We propose the following  principle:
\begin{principle} 
Any criterion for the resource status of a state that rests on its 
support for a set of MBQCs must require that set 
to be \imb. 
\label{resStatus}
\end{principle} 
We highlight that in this principle, as in all such general 
propositions in this paper, it is necessary to consider a {\em set} of MBQCs, 
not merely a single MBQC.
In keeping with Principle \ref{resStatus}, we also propose:
\begin{principle} 
A state $\rho$ is a {\it resource for MBQC} only if
it supports a set of computations that is \imb.
\label{resPrinciple}
\end{principle} 
As a consequence of Principle \ref{resStatus},
to establish a state as a resource, it does not suffice to 
identify a set of MBQCs it supports unless they are \imb.

The above principles do not involve any consideration of how $\rho$
may be generated. This is quite deliberate.
Whether a state can be efficiently generated is an interesting question,
but one that should be kept separate from whether a state
(even one that cannot be efficiently generated) should be considered a
resource for MBQC. Maintaining this separation allows for discussion of
distinctions between states in terms of their power for
MBQC regardless of whether those states are practical. This stance
is in keeping with discussions of universal resource states for MBQC.
For example, 
Refs.~\cite{Gross09,Bremner09}
consider the question of whether certain states are too entangled
to be resources for MBQC without considering whether they can be efficiently
generated. The question of whether there are states that can be
efficiently generated that are too entangled was addressed separately
\cite{Low09}.
More generally, when considering whether a state is
a resource for any task, it should not matter where that state came from.
In the current situation, for example, it should not matter
whether a putative resource state for MBQC was generated via MBQC or not.

We now turn to the question of  which computations are \imb. 

\section{Some simple criteria}
\label{sec:simpleCriteria}

As we noted above, there are different types of MBQC, with 
differing restrictions on the types of measurements and
the types of classical computation that can be done, and still more
types likely to be defined. 
For this reason, we do not give a full definition with necessary and 
sufficient conditions for a set of MBQCs to be \imb.
We give a series of increasingly stringent requirements for a set of 
MBQCs to be \imb, expressed as a series of increasingly 
weak sufficient conditions for a set of MBQCs to be \osmb, 
and therefore not \imb.
Informally, all of these conditions say that for 
a quantum computation to be considered \imb, 
measurements must form a key part of the computation.

Consider a set $S$ of MBQCs employing $\rho$, 
a pre-measurement state of $\nu$ qubits. 
We propose: 
\begin{criterion}
A set $S$ of MBQCs is \osmb\
if, for every computation in $S$, the measurement of the 
pre-measurement state 
always yields the same classical $\nu$-bit string $\boldsymbol{m}$. 
\label{crit1}
\end{criterion}

The reasoning here is that the quantum state $\rho$ can be replaced by 
the more conveniently stored classical bit string 
$\boldsymbol{m} = m_\nu\dots m_1$ 
with no loss whatsoever. In fact,
doing so saves the ultimate user the trouble of doing the measurements. 
Instead of performing a MBQC, the user 
performs only the classical
computational part of the MBQC, substituting in the appropriate bit values
from the classical bit string instead of performing the measurements.
The resulting output is
indistinguishable from that of the MBQC.
In such a computation, measurements are no longer needed, and 
the entire computation becomes completely classical.
It seems fair to say that such a computation is not \imb\
or quantum.

This criterion, together with Principle \ref{resStatus}, 
allow us to make the following uncontroversial 
statement. 
On the basis of a set of MBQCs that 
always measure the state $\rho$ in the computational basis, one cannot 
claim a state $\rho = \ket{\boldsymbol{m}} \bra{\boldsymbol{m}}$, 
corresponding to the classical bit string 
$\boldsymbol{m} = m_\nu\dots m_1$, 
to be a resource for MBQC. Nobody has, to our knowledge, 
made precisely such a claim, but this is an important step 
towards our more substantial conclusions later in this section.

We now propose a stronger criterion. 

\begin{criterion}
A set $S$ of MBQCs is \osmb\
if the  $\nu$-qubit pre-measurement state $\rho$ can be 
measured locally (i.e., each qubit separately) ahead of time,
without regard to which computation $f$ in $S$ will be carried out, 
and the resulting classical $\nu$-bit string $\boldsymbol{m}$ 
can be used to carry out any one of the computations in $S$.
\label{crit2}
\end{criterion}

Again, the reasoning here is that the quantum state $\rho$ can be 
replaced by a more
conveniently stored classical bit string with no loss whatsoever, 
saving the ultimate user the trouble of doing the measurements,
 and turning the computation into a completely classical one. 
 This criterion means that 
a set $S$ of MBQCs in which 
the measurement basis for each qubit 
of $\rho$ does not vary between
elements of $S$ should not be considered \imb. 
We stress that the classical bit string $\boldsymbol{m}$ here is not 
required to be the same in every  run of the algorithm. 
That is, the bit string $\boldsymbol{m}$ replaces the 
pre-measurement state $\rho$ 
in the sense that it can be used to yield (through purely classical 
processing, of course) a single output, exactly the same way 
that, in general, the pre-measurement state $\rho$ only 
allows a MBQC to be run once. To state this 
more formally, Criterion \ref{crit2} can be rephrased as follows: 
A set $S$ of MBQCs is \osmb\
if, given a set of $r$ copies of the pre-measurement state $\rho$, 
each can 
be measured locally ahead of time, without regard to which computations $f$
in $S$ will be carried out, and the resulting classical bit strings
$\boldsymbol{m}_1, \boldsymbol{m}_2, \dots, \boldsymbol{m}_r$ 
can be used to carry out any set of $r$ computations in $S$. 

It is critical to recognize that Criterion \ref{crit2} does not imply
that depth-1, or non-adaptive, MBQCs are \osmb. A set $S$ of 
non-adaptive MBQCs in which different members of the set $S$ require
the qubits to be measured in different bases avoids being classified
as \osmb\ by Criterion \ref{crit2} since the measurements cannot be
made ahead of time, without regard to the specific $f\in S$ 
to be carried out. Criterion \ref{crit2} only rules 
out any set $S$ of MBQCs in which the qubits are
measured in the same basis no matter what computation $f\in S$
is being carried out. That is, a distinction must be made 
between what we term {\it flexible} 
measurements of the initial state, in which the basis in which the qubits
are measured depends on the specific computation to be performed, 
and adaptive measurements, in which the basis in which some of the qubits
are measured depends on the outcome of previous measurements.
Criterion \ref{crit2} does not require adaptive measurements, 
only flexible measurements, and for this reason (as we
explain carefully in the next section) it does not imply that the MBQCs for 
{\it IQP} and {\it IQP}$^*$ starting with graph states are \osmb.

\section{Discussion of Hoban \ea's claims}
\label{sec:discOfHoban}

In Ref.~\cite{Hoban13}, Hoban \ea\ claim to ``show that there exist 
computations which cannot be efficiently and exactly performed on a 
classical computer, but can be performed in standard measurement-based 
quantum computation (MBQC) using resource states with zero entanglement 
and zero discord.'' In this section, we examine their arguments and
explain why they do not support this conclusion.
In Sec.~\ref{sec:genArg},
we present a more general argument that shows that discord-free states
cannot be resources for MBQC. 

The MBQCs on which Hoban \ea\ base their claim 
that discord-free states can be resources for 
MBQC involve measurement in only the eigenbasis 
of the discord-free state $\rho$. 
Thus, Criterion \ref{crit2} is strong enough to establish 
that these sets are not
\imb, and therefore, by Principle \ref{resStatus}, cannot be used 
to establish that a type of quantum state is a resource for MBQC. 

Before presenting our argument in more detail, 
it is worth mentioning that we have no disagreement 
with the other main claim of Hoban \ea~\cite{Hoban13}, 
namely their Lemma 1, 
in which they have applied similar proof techniques to 
those in \cite{Bremner11} to show that efficient classical sampling of the 
output of {\it IQP}$^*$ circuits, a subclass of {\it IQP} circuits
with a more standard uniformity condition, implies
the collapse of the polynomial hierarchy to the third level, just
as it does for {\it IQP} circuits \cite{Bremner11}. 
Other sampling problems that can be achieved efficiently using quantum means
and are unlikely to be efficiently achievable classically are known
\cite{Aaronson11, Aaronson13, Terhal04}. 

First, we review the construction of Hoban \ea~\cite{Hoban13}, 
so that we can discuss issues specific 
to their arguments. They define an {\it IQP}$^*$ 
circuit family to be a family
of quantum circuits $C_{Z,\Theta}$, indexed by the integers $n$, 
acting on a register of $\nu$ qubits 
and taking as input an $n$-bit string $\boldsymbol{x} = x_n\dots x_1$. 
The family must be uniform, meaning that there exists a Turing 
machine that can output a description of the quantum circuit in polynomial time 
given input $n$. For {\it IQP}$^*$, that description specifies 
(i) how $\nu > n$ depends polynomially on $n$;
(ii) $Z$, a set of polynomially-many $\nu$-bit strings 
$\boldsymbol{z} = z_\nu\cdots z_1$; and
(iii) $\Theta$, a set of angles $\theta_{\boldsymbol{z}}$, one 
for each string $\boldsymbol{z}\in Z$, such that each 
$\theta_{\boldsymbol{z}} \in (0,2\pi]$ has a polynomial-sized description. 
The circuit consists of a polynomial sequence of gates of the form
$e^{\i\theta_{\boldsymbol{z}} X [\boldsymbol{z}]}$, 
where $X[\boldsymbol{z}] = \bigotimes_{j=1}^{\nu}X^{z_j}$. The 
register is initialized to the $n$-qubit state 
$\ket{\boldsymbol{x}}$ followed by $(\nu-n)$ qubits initialized to $\ket{0}$. 
The $\nu$-bit output $\boldsymbol{m}$
is obtained by measuring all of the qubits in the
computational basis at the end of the computation.

Hoban \ea\ prove that efficiently and exactly simulating the output of
such circuits classically would imply the collapse of the polynomial 
hierarchy to the third level. They have thereby proven
their first result, Lemma 1 -- an interesting result, 
with which, as we said, we have no quarrel. 

They further show that all
of the computational difficulty in simulating the output of these circuits
classically is in the simulation of the output for any one particular 
input $\boldsymbol{{x}}$. For example, one can restrict 
to the $\boldsymbol{0}$ input state in which the input is a 
string of $n$ zeros, 
as can be seen from the following symmetry argument. 
Let $\bar{\boldsymbol{x}}$ be the $\nu$-bit string consisting of 
$\boldsymbol{x}$ followed by $\nu-n$ zeros. 
The probability $P(\boldsymbol{m}|\boldsymbol{x})$ that 
output string $\boldsymbol{m}$ is obtained, 
given input string $\boldsymbol{x}$, is equal to 
$P(\boldsymbol{m}\oplus\bar{\boldsymbol{x}}|\boldsymbol{0})$. 
Thus, given a way to obtain samples for the probability distribution 
for input $\boldsymbol{0}$,
samples for the probability distribution for input $\boldsymbol{x}$ can be 
obtained trivially. 

Hoban \ea\ exhibit a way to implement the circuits $C_{Z,\Theta}$
acting on input $\boldsymbol{x}= \boldsymbol{0}$ using MBQC. The MBQC
uses a graph state $\rhodf_Z$ with $r  = \nu + |Z|$ qubits 
where $|Z|$ is the number of
elements $\boldsymbol{z}\in Z$ or, equivalently, the number of gates applied 
in the original circuit. We will use 
$c_j$, with $j \in \{1,\cdots, \nu\}$ 
for the computational qubits, and 
$q_{\boldsymbol{z}}$ for the additional $|Z|$ qubits. Edges go from a qubit 
$q_{\boldsymbol{z}}$ to
exactly the subset of the $\nu$ qubits $c_j$ such that the $j$th 
bit of $\boldsymbol{z}$, $z_j$, is $1$. 
To perform the computation, 
each of the $|Z|$ qubits $q_{\boldsymbol{z}}$ is measured in the basis 
corresponding to its associated angle $\theta_{\boldsymbol{z}}$
yielding $|Z|$ bit values $b_{\boldsymbol{z}}$. 
To complete the computation, the remaining $\nu$ qubits $c_j$ are all 
measured in the $X$-basis $\{\ket{+},\ket{-}\}$, and on the resulting bit
values a classical computation determined by the $\{b_{\boldsymbol{z}}\}$
is performed to obtain the $\nu$-bit outcome $\boldsymbol{m}$. 
To obtain results for a circuit $C_{Z,\Theta}$ acting on
input $\boldsymbol{x} \ne \boldsymbol{0}$, one 
can simply (classically) compute
$\boldsymbol{m}\oplus\bar{\boldsymbol{x}}$. 
The full computations just described are undoubtably true MBQCs.
Each graph state $\rhodf_Z$ can support any computation 
in {\it IQP}$^*$ with the same set $Z$ but differing $\theta_{\boldsymbol{z}}$.
For this reason, they are {\em not} found to be 
\osmb\ according to our Criteria 
\ref{crit1}, \ref{crit2}, and \ref{crit3}. 
Indeed, this construction provides a measurement-based quantum 
computational means to efficiently sample a distribution that
is strongly suspected not to be able to be sampled efficiently classically.

Hoban \ea\ then point out that, starting with the graph state $\rho_Z$, one can  
dephase all of the qubits, the $q_{\boldsymbol{z}}$ 
in the $\theta_{\boldsymbol{z}}$ basis and the 
$c_j$ in the $X$ basis, to 
obtain the discord-free \cite{OllZur01,HenVed01} state  
\beq \label{zerodisc}
\rhodf_{Z,\Theta} = \sum_{\boldsymbol{y}=0}^{2^r -1 }P(\boldsymbol{y}|\boldsymbol{x})
\ket{\boldsymbol{y}}\bra{\boldsymbol{y}}
\eeq 
where here $\ket{\boldsymbol{y}}=\ket{y_r}\cdots\ket{y_1}$, and 
$\boldsymbol{y}$ is the bit string $y_r \cdots y_1$. 
That is, the state is ``essentially a classical probability distribution''  
\cite{Hoban13} and obviously is unentangled.  
Measuring each of the qubits in the basis
in which it was dephased, and then performing a classical computation
on the resultant bits -- exactly the
computation that would have been performed on the measurement outcomes
in the original MBQC -- 
yields output $\boldsymbol{m}$, 
a sample from the distribution $P(\boldsymbol{m}|\boldsymbol{x})$  
corresponding to the original circuit $C_{Z, \Theta}$.
Note that the state in \erf{zerodisc} is specific to 
the circuit $C_{Z, \Theta}$, 
which we have emphasized by denoting it $\rhodf_{Z, \Theta}$.  
Thus, measuring $\rhodf_{Z, \Theta}$ yields samples from only this one
distribution, though bit-wise addition of $\bar{\boldsymbol{x}}$
enables sampling from the trivially related 
distributions obtained by applying $C_{Z, \Theta}$ 
to input $\boldsymbol{x} \ne \boldsymbol{0}$.

We fully agree with the correctness of these statements, but we disagree 
markedly with Hoban \ea\ in the interpretation of this procedure. 
They claim that $\rhodf_{Z, \Theta}$ is a resource for MBQC.  
However, it is clear that there is no reason to stop their 
procedure after creating the discord-free state $\rhodf_{Z, \Theta}$.
One can simply measure it and obtain a classical bit string 
$\boldsymbol{y}$ that can be stored classically without 
requiring quantum storage. 
Indeed, any computation that uses a quantum
state as a resource only to  measure each of its
qubits in a fixed basis 
could be performed identically and more conveniently 
using a classically stored bit string. 
Thus, by Criterion \ref{crit2} and Principle \ref{resStatus}, 
it is clear that this set 
of of computations does not support Hoban \ea's claim that $\rhodf$ 
is a resource for MBQC. 

It is important to understand why Criterion \ref{crit2} implies
that the computations starting with a discord-free state $\rho_{Z,\Theta}$
are \osmb, while the computations starting with a graph state $\rho_Z$ 
are not so categorized.  A single initial graph state $\rho_Z$ can support a 
set of computations
$S_Z = \{f_\Theta\}$ for all possible sets of measurement angles $\Theta$.
Criterion \ref{crit2} does not classify
$S_Z$ as \osmb\ because $\rho_Z$ cannot
be measured ahead of time and still support the computation of all
$f_\Theta\in S_Z$; the measurements to be performed depend on which
$f_\Theta$ is to be carried out. In the language of 
Sec.~\ref{sec:simpleCriteria}, while
all of the MBQCs in this set are non-adaptive, the set does 
require flexible measurements. 

On the other hand, a single discord-free state $\rho_{Z,\Theta}$ 
obtained by dephasing as described above, supports a much smaller
set of the above computations than the graph state $\rho_Z$. It supports only
the set of above computations in which each of the qubits $q_z$ is 
measured in the basis associated with $\theta_z$, where $\Theta$ contains the 
angles $\theta_z$, one for each qubit $q_z$. For this set of computations, 
flexible measurements are not required; one can still perform all of the
computations in the set after measuring in the eigenbasis of $\rho_Z$. 
One does not need to know which computation is desired before measuring
all of the qubits. Thus, this smaller set of computations is \osmb\ by
Criterion \ref{crit2}.

One can gain further insight by recognizing that Hoban \ea's construction 
of an MBQC to obtain their discord-free state 
is completely unnecessary: they could just as
well have dephased the state obtained in their original quantum circuit
model prior to the final measurement of all of the qubits $c_j$ 
in the standard 
(logical) basis. This would yield the even simpler discord-free state 
\beq \label{zerodisc-m}
\rho_{Z,\Theta}' = \sum_{\boldsymbol{m}=0}^{2^\nu -1 }P(\boldsymbol{m})\ket{\boldsymbol{m}}\bra{\boldsymbol{m}}.
\eeq 
where $P(\boldsymbol{m}) = P(\boldsymbol{m}|\boldsymbol{x})$ is the desired 
probability distribution for the  ``measurement output string \ldots $\boldsymbol{m}$'' \cite{Hoban13}. 
In deciding whether or not a state should be considered a resource state
for MBQC, it should not matter how the state was obtained. In particular,
it should not matter whether it was obtained through MBQC or through some
other means. 

The same construction can be applied to many quantum computations. 
In the following subsection, we consider 
Shor's factoring algorithm \cite{Shor-94,RPbook} as an example. 

\subsection{An analogous ``resource'' for factoring} 

Suppose $F$ is a 
uniform family of integers $M_n$, by which we mean that a Turing 
machine can output $M_n$, an integer of bit length $n$, in polynomial time 
given input $n$. For example, we may use a cryptographic grade pseudorandom
number generator to output a series of integers of increasing length. 
The expected runtime of any known classical factoring algorithm
is superpolynomial in $n$, and it is strongly suspected 
that any classical
approach to factoring is superpolynomial. For this reason, most of the 
numbers in these uniform families will be hard to factor classically. 

Next we introduce, as is standard for Shor's algorithm, 
another efficient pseudorandom number generator $G$ that
outputs another uniform family of integers $a_n < M_n$.
We define uniform Shor circuit families to be families in which each circuit
is a concatenation of the following: (i) a quantum (though 
it may as well be classical) circuit that upon input $n$ computes
$M_n$ and $a_n$; (ii) a quantum circuit that takes as input $M_n$ and 
$a_n$ and a register of size $\nu + n$, where $M_n^2 \leq 2^\nu < 2M_n^2$, 
prepared as an equal superposition of all logical 
values $x$ for $0\leq x \leq 2^{\nu}-1$ in the first $\nu$ bits of 
the register, and computes the function $f_n(x) = a_n^x \mod M_n$ 
in a quantum parallel sense into the remaining $n$ bits of the register; 
and (iii) a quantum circuit of sufficient 
size (polynomial in $n$) that it can carry out Shor's algorithm 
for this input. Finally, a read-out is, of course, performed to obtain, 
nondeterministically,  
an estimate of the period of $f_n$, which, with high probability, 
provides a key to factorizing $M_n$. 

Now, instead of performing this  final measurement, one 
could dephase the final quantum state in the computational basis,
obtaining a discord-free state $\rho$. Since, with high probability, a
factor of $M_n$ can be obtained from a sample of the classical probability
distribution associated with $\rho$, such probability distributions 
in general cannot be sampled efficiently classically given 
input $n$ and the description of the circuit, unless there is a classical
factoring algorithm of expected polynomial runtime.

Should such a $\rho$ be considered a resource for MBQC? 
Just as we do not believe that Hoban \ea's arguments support their Theorem 2, 
we do not believe that the analogous argument we have just given 
supports the analogous theorem,  
{\it Shor's algorithm may be implemented in MBQC using 
resource states with zero entanglement and zero discord.} 
Similarly, 
we do not believe such arguments support their Corollary 2 or its analog, 
{\it Shor's algorithm may be implemented in MBQC using correlations which do not
violate any Bell inequalities.} 
Criterion \ref{crit2} together with
Principle \ref{resStatus} explain why.  

We now turn to showing why discord-free states could never be resources
for MBQC.

\section{General argument}
\label{sec:genArg}

In this section, we give a third, broader criterion that encompasses additional
sets of MBQCs that, we argue, should be considered \osmb.
We use this third
criterion to establish that discord-free states cannot 
be considered resources for MBQC. 
Before motivating the broader criterion we will define, we first recall 
the definition of a discord-free state \cite{OllZur01,HenVed01}. 
For $\nu$ qubits, it is any state 
that can be written, 
for a suitable definition of basis $\{ \ket{0},\ket{1}\}$ for each qubit, 
in the form of \erf{zerodisc-m}, 
where $P(\boldsymbol{m})$ is an arbitrary probability distribution. 

A state of this form is not only 
``essentially a classical probability distribution'' $P(\boldsymbol{m})$ 
\cite{Hoban13}; it is experimentally indistinguishable from a 
 pre-existing basis state $\ket{\boldsymbol{m}}$ with the 
bit string $\boldsymbol{m}$ drawn at random  from that distribution 
\cite{WisemanMilburn}. As an obvious consequence,  
any computation that can be done with a 
discord-free state  $\rho$ is indistinguishable 
from one done with a sample $\ket{\boldsymbol{m}}$ from 
the probability distribution represented by the discord-free state $\rho$. 

Suppose one starts with a sample $\ket{\boldsymbol{m}}$ from 
the probability distribution represented by the discord-free state $\rho$. 
Instead of making the single-qubit measurements on 
state $\ket{\boldsymbol{m}}$ (according to whatever putative 
MBQC algorithm is to be implemented), one could 
generate results with the same statistics as these 
measurement results using the classical 
bit string $\boldsymbol{m}$ 
and classical computation on each of the $\nu$ bits of $\boldsymbol{m}$ 
individually, independent of the values of any of the other bits.
Specifically, for each qubit $j$,
given $m_j$ and the measurement basis, one can trivially 
calculate the probability distribution 
from which the output bit may be generated. 

Note that a single discord-free state can be measured only once, 
so a discord-free state is
not a black box that provides samples from a distribution, but
rather one that yields only a single sample. Perhaps it will help
the reader to understand if we restate this in the style of the discussion 
following Criterion~\ref{crit2}. A set of $100$ discord-free
states $\rho$, all of which are measured in the eigenbasis of $\rho$, 
can be replaced
by a set of $100$ samples from the distribution. Thus, a set
of discord-free states can be replaced by a bit string 
containing the same number of samples as the number of 
discord-free states one is replacing. 

These considerations motivate
\begin{criterion}
A set $S$ of MBQCs should be considered \osmb\
if the $\nu$-qubit state $\rho$ can be measured locally ahead of time,
without regard to which computation $f$ in $S$ will be carried out, 
and the resulting classical $\nu$-bit string $\boldsymbol{m}$ can be used, 
together with classical (possibly non-deterministic) computation on each bit 
of $\boldsymbol{m}$ separately,
to carry out any one of the computations in $S$.
\label{crit3}
\end{criterion}

For the same reasons explained with respect to Criterion \ref{crit2}, 
Criterion \ref{crit3} does not imply that the MBQCs for 
{\it IQP} and {\it IQP}$^*$ starting with graph states are \osmb.
The paragraph prior to stating Criterion \ref{crit3} shows that
any set of MBQCs supported by a discord-free state satisfies
Criterion \ref{crit3} and is therefore not \imb. This statement,
together with Principle \ref{resStatus}, implies that
discord-free states cannot be resources for MBQC.
This completes our argument for why 
it would never make sense to claim that a discord-free 
state could be a resource for MBQC.

As a final note, we remark that more stringent 
conditions than the ones we have given are likely called for.
It seems reasonable, for example, that if, in a set of MBQCs, 
the pre-measurement state could be replaced by a 
classical bit string with only 
linear classical computational overhead, then said set is not \imb.
Some readers may worry that such a principle 
would rule out graph states as a resource for MBQC, because 
in many cases graph states can be generated from a
classical description using only linear computation. However, 
the computation required is quantum, not classical. 
Thus, while a graph state may be a resource state for MBQC,  
its classical description is not; nor do  Refs.~\cite{WisemanMilburn, RPbook}, 
however useful, constitute resources for quantum computation.
Other readers may wonder why we restrict the
amount of classical computation.  If we were to allow  unlimited 
classical computation, all measurement-based
computation, and in fact, all quantum computation, could be carried out
since {\it BQP} is in {\it PSPACE}. For this reason, limiting the amount of
classical computation is necessary.\\ \\

\section{Conclusion} \label{sec:conc}

We have proposed the principle that a state
should be considered a resource for MBQC only if it supports
\imbh\ quantum computations. 
We have also provided necessary criteria for MBQCs to be \imb, 
criteria which are so weak as to be, we hope, uncontroversial. 
Building from this principle and these criteria, we have
explained why it could never be correct to consider discord-free states 
as resources for MBQC, and we have specifically addressed the arguments
of Hoban \ea\ that suggested otherwise. 

We leave the problem of fully characterizing what it means for 
a state to be a resource state for MBQC, and what it means for a 
MBQC to be \imb, as future work. 

{\em Note added:} 
Subsequent to the the submission of our paper, Hoban \ea\ have revised 
their paper, removing the claim that discord-free states can be 
resources for MBQC 
and changing their title \cite{Hoban13v3}. Nevertheless, 
their arguments remain much as they were, so most of the detailed critique 
we give in Sec.~\ref{sec:discOfHoban} applies just as much to their 
revised version as to the original. 
In particular, our criticism in Sec.~\ref{sec:discOfHoban} 
still holds: the class of problems solved by any
one of their discord-free “resource states” is a trivial class
quite distinct from {\it IQP}$^*$; moreover, their dephasing procedure
has nothing to do with {\it IQP}$^*$, but rather could be applied to
any quantum algorithm (including Shor's, as we discuss) to 
produce a “resource” of similarly limited utility.

\section*{Acknowledgements}
Both authors would like to thank Dan Browne for stimulating e-mail
exchanges.

\bibliography{qc}

\end{document}

%% file: defs.tex
\def\ket#1{|{#1}\rangle}
\def\bra#1{\langle{#1}|}

\def\i{\mathbf {i}}

\def\mod{\mathop{\mbox{mod}}}

\newtheorem{theorem}{\noindent{\bf Theorem}}[section]

